\documentclass[aps]{revtex4}
\usepackage{graphicx}
\DeclareGraphicsExtensions{.pdf,.eps}

\newcommand{\srr}[1]{\stackrel{\rightharpoonup\!\!\!\!\rightharpoonup }{#1}}
\newcommand{\Tep}[0]{\srr{\cal\varepsilon}}
\newcommand{\Tepmain}[0]{\Tep_{\!\!main}}
\newcommand{\Tup}[0]{\srr\Upsilon}
\newcommand{\Ttau}[0]{\srr{\cal\tau}}
\newcommand{\Tpi}[0]{\srr{\cal\pi}}
\newcommand{\TI}[0]{\srr I}
\newcommand{\mbf}[0]{\mathbf f}
\newcommand{\mbF}[0]{\mathbf F}
\newcommand{\bh}[0]{\mathbf h}
\newcommand{\br}[0]{\mathbf r}
\newcommand{\sh}[0]{\mathbf\sigma_{\bh}}
\newcommand{\bT}[0]{\mathbf T}
\newcommand{\bF}[0]{\mathbf F}
\newcommand{\bef}[0]{\mathbf f}
\newcommand{\ba}{\mathbf a}
\newcommand{\bb}{\mathbf b}
\newcommand{\bc}{\mathbf c}
\newcommand{\bhz}[0]{\mathbf h_{0}}
\begin{document}
\title{Dynamical Equations For The Period Vectors In A Periodic System Under Constant External Stress}
\author{Gang Liu}
\address{\\
{\small e-mail: gang.liu@queensu.ca}\\
{\small High Performance Computing Virtual Laboratory}\\
{\small Queen's University, Kingston, Ontario, Canada}\\
%{\small Queen's University, Kingston, Ontario, K7L 1H3, Canada}\\
%{\small 115-993 Princess Street}\\
%{\small Kingston, Ontario, K7L 1H3, Canada}\\
%{\small fax: (613) 533-2015}\\
%{\small tel: (613) 533-6000 ext.78387}\\
%{\small e-mail: gang.liu@queensu.ca}\\
}

\date{Revised on December 16, 2014}

%\pacs{61.50.Ah, 62.50.-p, 63.20.dk, 64.10.+h.}
%
%  pacs: 02.70.Ns   Molecular dynamics and particle methods
%
%  pacs: 45.05.+x   General theory of classical mechanics of discrete systems
%
%  pacs: 45.20.-d    Formalisms in classical mechanics
%
%  pacs: 45.20.D-   Newtonian mechanics
%
%  pacs: 47.10.-g    General theory in fluid dynamics
%
%  pacs: 61.50.Ah   Theory of crystal structure, crystal symmetry; calculations and modeling
%
%  pacs: 62.50.-p    High-pressure effects in solids and liquids 
%
%  pacs: 63.20.dk   First-principles theory
%
%  pacs: 64.10.+h  General theory of equations of state and phase equilibria 
%

\maketitle

%\newpage

%\begin{abstract}
\section*{Abstract}
The purpose of this paper is to derive the dynamical equations for the period vectors of a periodic system under constant external stress. The explicit starting point is Newton's second law applied to halves of the system. Later statistics over indistinguishable translated states and forces associated with transport of momentum are applied to the resulting dynamical equations. In the final expressions, the period vectors are driven by the imbalance between internal and external stresses. The internal stress is shown to have both full interaction and kinetic-energy terms.
%\end{abstract}
%\maketitle

%\newpage

\section{Introduction}

As an application of Newtonian Dynamics, Molecular Dynamics (MD) simulations of many-particle systems play an important role in Biophysics, Chemical Physics, and Condensed Matter Physics\cite{al,fre,haile}. As simulated systems, crystals are of periodic structures by themselves. For other systems, the periodic boundary conditions are often employed to reduce the number of degrees of freedom. In these situations, the system being studied can be described as a ``dynamical crystal" filled with repeating cells.  The particle position vectors inside a given cell,  together with the independent period vectors (cell edge vectors, or basic vectors or lattice vectors in crystals) form a complete set of degrees of freedom of the system. The dynamics of the particles in the cell are determined by Newton's second law. What about the period vectors?

In 1980, Andersen proposed an MD theory that allows the cell volume (but only the volume) to change in a simulation of fluids, subject to a constant external pressure \cite{a}. Shortly thereafter, Parrinello and Rahman extended Andersen's idea to a new MD theory (PRMD), in which the dynamical equations for the complete period vectors were proposed for constant external pressure and constant external stress respectively\cite{pr1,pr2}, both in the framework of Lagrangian Dynamics. 

Since then, PRMD has been extensively used in simulation studies, for instance for phase transitions of crystal structures induced by external forces\cite{Mart,Rou,mol,Mor,Tra}. PRMD was also combined with the well-known Car-Parrinello MD \cite{cpmd,fo,ber}. Several other versions of PRMD under constant external pressure have since been proposed \cite{ray1,ray2,ray3,cll,wen,jv,szm}.  These indicate that the dynamics of the period vectors is an important physics problem that had not been solved before.

However, in 1983, Nos\'{e} and Klein pointed out that ``The result is the usual Newton's second law equation with a correction term
arising from the change in shape of the MD cell" (just before Eqn (2.5) on page 1057 in \cite{nose}). This implies that the dynamics of the particles generated in PRMD is not Newton's second law, due to the incomplete kinetic energy of particles in their Lagrangian.  Another drawback in PRMD is that the generated dynamical equations of the period vectors under constant external stress do not have a form where the period vectors are driven by the imbalance between internal and external stresses (see Eqn (2.25) in \cite{pr2}). In any case, PRMD can find the true equilibrium states under zero temperature and constant external pressure, when all velocities and accelerations are zero.

In 1997, inspired by the ideas in PRMD, we derived another set of dynamical equations for the period vectors in the case of constant external pressure, starting from a Lagrangian that includes the full kinetic energy of particles. However, we did not obtain a full expression of the internal stress in these dynamical equations, while the generated  dynamics for the particles is Newton's second law\cite{lww,lwssc}. 

Since all particles in all cells have to obey Newton's second law, which will be kept intact throughout this paper, the complete dynamical equations of the period vectors must be derivable in the framework of Newtonian Dynamics directly. The purpose of this paper is to perform such a task specifically. As a first step, we apply Newton's second law to halves of the whole system while considering external stress explicitly. Then statistics over indistinguishable  translated states and forces associated with transport of momentum are applied to the obtained dynamical equations. In the resulting expressions, the period vectors are driven by the imbalance between internal and external stresses, so that in the corresponding equilibrium states, the two stresses balance each other. Furthermore, the internal stress is shown to have both full interaction and kinetic-energy terms.

This paper is organized as follows. The model is given in Sec. \ref{sec:Model}. 
The application of Newtonian Dynamics on half systems is presented in Sec. \ref{sec:Instancedynamics}. 
The statistics over indistinguishable translated states and forces associated with transport of momentum to the obtained dynamical equations are done in Sec. \ref{sec:Statistics}. 
Sec. \ref{sec:Summary} is devoted to summary and discussion.

\section{Model}\label{sec:Model}

The macroscopic bulk of a material with an inside microscopic periodic structure is taken as the model for this study. 
We use $\mathbf{a}$, $\mathbf{b}$, and $\mathbf{c}$ as the three independent period vectors, forming a right-handed triad. Then each cell can be denoted by the corresponding lattice translation vector 
$\mathbf{T} = T_\mathbf{a} \mathbf{a} + T_\mathbf{b} \mathbf{b} + T_\mathbf{c} \mathbf{c}$,  
with integers $T_\mathbf{a}$, $T_\mathbf{b}$, $T_\mathbf{c}$ ranging from negative infinity to positive infinity.  As usual,  let us call the centre cell of ${\mathbf T}=0$ the ``MD cell'', and the particles in that cell the ``MD particles". Since we study the properties of the inner part of the bulk around the MD cell, far-away surface effects are neglected. Then again, as in PRMD, all vectors of the MD particle positions and the periods are the full degrees of freedom of the system.

The dynamics for the MD particles is described by Newton's second law  
\begin{equation}
m_i \ddot \mathbf{r}_i = \mathbf{F}_i\ \ (i=1,2,\cdots ,n),\label{b03}
\end{equation}
where $\mathbf{r}_i$ is the position vector of the $i$th MD particle with mass $m_i$. $\mathbf{F}_i$ is the net force acting on MD particle $i$ from all other particles of any cell, but no external force due to distance, and $n$ is the total number of MD particles. Then the only unknown dynamical equations are the ones for the period vectors. 

The external action on the surface is expressed by the constant external stress tensor (or dyad) $\Tup$ . 
The corresponding external forces are modeled as applied by the surrounding external walls contacting the surface of the bulk.
For the case of constant external pressure $p$ , 
$\Tup=p\TI$, 
where $\TI$ is an identity tensor or unit matrix, and the positive direction is defined from inside to outside of the bulk. By definition, the external force acting on an infinitesimal surface area vector $d{\mathbf s}$ of the bulk is $d\mathbf{F} = \Tup\cdot\, d\mathbf{s}$.
The net external force on the bulk is 
\begin{equation}
\mathbf{F} = \oint_{sf}\Tup\cdot\, d\mathbf{s} = 
\Tup\cdot\,\oint_{sf}d\mathbf{s}=0,  \label{b02}
\end{equation}
where the integral is over all the surface of the bulk, and therefore no acceleration for the bulk as a whole. The external stress $\Tup$ is assumed to be symmetric, i.e., for all of its components $\Upsilon _{i,j}=\Upsilon _{j,i}$. 
This  assumption ensures that the net external torque on the bulk is zero. We further assume that the bulk does not rotate. 

In this paper, we only consider interactions of the form of pair potentials 
$\varphi^{(2)}(r_{i,j})=\varphi^{(2)} (\left| {\mathbf r}_i-{\mathbf r}_j\right| )$ 
between particles. Extension to many-body interactions can be done by refering our previous papers\cite{lww,lwssc}.
From Newton's Third Law and the periodicity of the system, it follows that the net of all forces acting on all MD particles is zero, i.e.
\begin{equation}
\sum_{i=1}^nm_i\ddot {{\mathbf r}}_i=\sum_{i=1}^n{\mathbf F}_i=0.  \label{b12}
\end{equation}
Let us then employ the centre-of-mass coordinate system of MD cell for all the work throughout this paper. 
Then the total momentum of the MD cell is
\begin{equation}
\sum_{i=1}^nm_i\dot {{\mathbf r}}_i=0.  \label{bmmt12}
\end{equation}

As the period vectors may change with time, the volume $\Omega = \left(\mathbf{a}\times\mathbf{b}\right) \cdot \mathbf{c}$ and shape of the MD cell and those of the bulk should also change correspondingly.

\section{instantaneous dynamics}\label{sec:Instancedynamics}

\begin{figure}
  \begin{center}
    \hspace{-0.5cm}
    \includegraphics[width=0.7\textwidth]{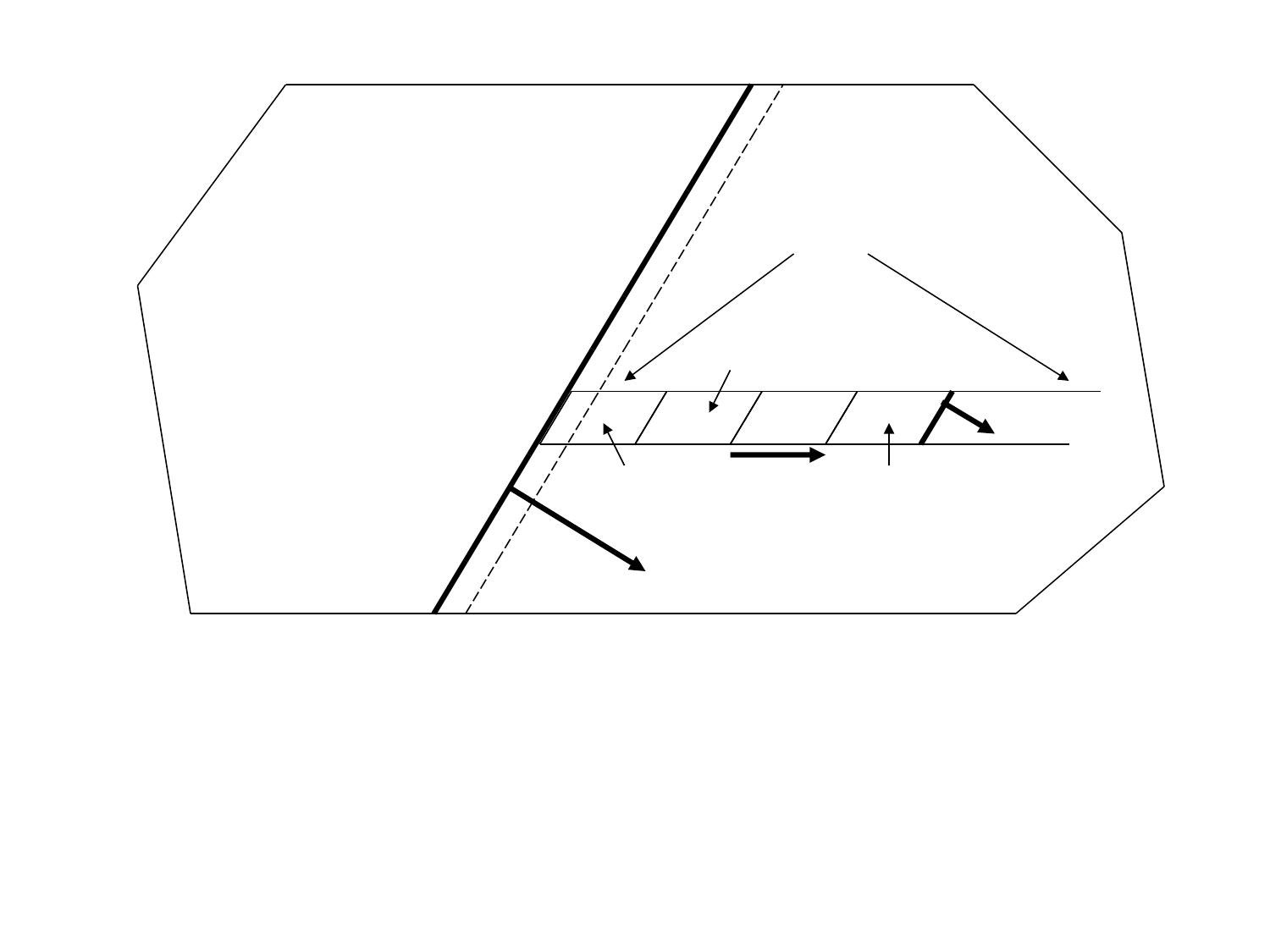}
  \end{center}

  \vspace{-9.5cm} \hspace{2.3cm} $P$ \hspace{0.05cm} $Q$ \\
  \vspace{1.2cm} \hspace{1.0cm} $L_{\mathbf h}$ \hspace{3.6cm}  $R_{\mathbf h}$ \\
  \vspace{0.0cm} \hspace{3.4cm} $B_{\mathbf h}$  \\
  \vspace{0.6cm} \hspace{1.3cm} cell ${\mathbf h}$  \\
  \vspace{0.2cm} \hspace{6.8cm} ${\mathbf\sigma}_{\mathbf h}$  \\
  \vspace{0.4cm} \hspace{2.0cm} MD cell \hspace{0.4cm} ${\mathbf h}$  \hspace{0.4cm} cell $3{\mathbf h}$\\
  \vspace{0.4cm} \hspace{-1.8cm} ${\mathbf S}$  \\
  \vspace{0.5cm} \hspace{-4.2cm} $P^\prime$ \hspace{0.05cm} $Q^\prime$ \\
  \vspace{7.3cm} 
  \vspace{-0.6cm} 
  \vspace{-0.1cm} 
  \vspace{-0.3cm} 
  \vspace{-0.2cm} 
  \vspace{-0.2cm} 
  \vspace{-0.1cm}
  \vspace{-0.3cm} 

  \vspace{-5.0cm}
  \caption{
A sketch for the bulk of a material of a periodic structure being cut by plane $PP^{\prime }$, 
with a cross section area vector $\mathbf S$. Plane $PP^{\prime }$ is 
chosen such that, for a given period vector 
$\mathbf h=\mathbf a$, $\mathbf b$, or $\mathbf c$, 
the right ($R_{\mathbf h}$) part contains all 
$\mathbf{T} = T_\mathbf{a} \mathbf{a} + T_\mathbf{b} \mathbf{b} + T_\mathbf{c} \mathbf{c}$
cells with $T_{\mathbf h}\geq 0$, and the left ($L_{\mathbf h}$) part contains all the 
rest  ${\mathbf T}$ cells with $T_{\mathbf h}< 0$.  
The ``half cell bar'' $B_{\mathbf h}$ is composed of the MD cell and cells $\mathbf h$, 
$2\mathbf h$, $3\mathbf h$, $4\mathbf h$, etc., till the surface.
Newton's second law is applied to the $R_{\mathbf h}$ part for the dynamical equations
of the period vectors.
}
  \label{fig1}
\end{figure}

In order to find the dynamical equations for the period vectors, let us imagine a plane $PP^{\prime }$ that cuts the model bulk into a right part and a left part, with ${\mathbf S}$ as the area vector of the cross section between the two parts in the direction of pointing to the right part, as shown in Fig.1. Plane $PP^{\prime }$ is chosen such that, for a given period vector 
$\mathbf{h} = \mathbf{a}$, $\mathbf{b}$, or $\mathbf{c}$, the right ($R_\mathbf{h}$) part contains all 
$\mathbf{T}=T_{\ba}{\ba} +T_{\bb}{\bb} +T_{\bc}{\bc}$
cells with $T_{\bh}\geq 0$, and the left ($L_{\bh}$) part contains all the 
rest  ${\mathbf T}$ cells with $T_{\bh}< 0$.

Now let us apply Newton's second law to a ``snapshot'' of the right ($R_{\bh}$) part. 
The net external force acting on the $R_{{\mathbf h}}$ part is 
\begin{equation}
\bF_{E,R} = \int_{R_{\bh},sf}\Tup\cdot
d{\mathbf s}=\Tup\cdot \int_{R_{\bh},sf} d{\mathbf s}=
\Tup\cdot {\mathbf S},\label{b04}
\end{equation}
where the integral is over the surface of the bulk in the $R_{\bh}$ part. Let ${\bF}_{L\rightarrow R}$ be the net force acting on the $R_{\bh}$ part by the $L_{\bh}$ part. Then the dynamical equation of the $R_{\bh}$ part is 
\begin{equation}
M_R\ddot {\br}_{RC} = {\bF}_{L\rightarrow R} + \Tup\cdot {\mathbf S},\label{b05}
\end{equation}
where $M_R$ and $\ddot {\br}_{RC}$ are the total mass and acceleration of the centre of mass of the $R_{\bh}$ part respectively. 

Since surface effects are neglected,  ${\bf F}_{L\rightarrow R}$ should be uniformly distributed cell by cell across the section ${\mathbf S}$ between the two parts. Let us  divide Eq.(\ref{b05}) by 
\begin{equation}
N_{\bh}=\left| {\mathbf S}\right| /\left|\sh\right|,\label{b07.0}
\end{equation}
where 
$\mathbf\sh=\partial\Omega /\partial {\bh}$ is the (right) surface area vector of a cell with respect to the period $\bh$, then 
\begin{equation}
\frac 1{N_{\bh}} M_R\ddot {\br}_{RC} = {\bF}_{\bh} + 
\Tup\cdot \sh,\label{b07}
\end{equation}
where ${\bF}_{\bh}$ is the net force, by the $L_{\bh}$ part, acting on the ``half cell bar'' $B_{\bh}$ composed of the MD cell and cells $\bh$, 
$2\bh$, $3\bh$, $4\bh$, etc., till the surface, as shown in Fig.1.

Denoting ${\bef}_{i,\bT\rightarrow j,\bT^\prime}$ as the force acting on particle $j$ of cell ${\mathbf T}^{\prime }$  from particle $i$ of cell ${\mathbf T}$, and considering the periodicity of the system as previously in \cite{lww,lwssc}, the net force, by the $L_{\bh}$ part, acting on cell $k\bh$ ($k \ge 0$) is 
\begin{eqnarray}
{\bF}_{{\bh}, k} &=& \ 
\sum_{\bT}^{(T_{\bh}<0)} 
\sum_{i,j=1}^n {\bef}_{j,\bT \rightarrow  i, k \mathbf h} = \ 
\sum_{\bT}^{(T_{\bh}<0)} 
\sum_{i,j=1}^n {\bef}_{j,\bT - k \bh \rightarrow  i, k \bh - k \bh} = \
\sum_{\bT}^{(T_{\bh}<0)} 
\sum_{i,j=1}^n {\bef}_{j,\bT - k \bh \rightarrow  i,\mathbf 0} \nonumber \\
&=& \ 
\sum_{\bT}^{(T_{\bh}<-k)} 
\sum_{i,j=1}^n {\bef}_{j,\bT \rightarrow  i,\mathbf 0} = \ 
\sum_{\bT}^{(T_{\bh}<-k)} 
\sum_{i,j=1}^n - {\bef}_{i,\mathbf 0\rightarrow j,\bT},
\end{eqnarray}
or in another view
\begin{eqnarray}
{\bF}_{{\bh}, k} &=& \ 
\sum_{\bT}^{(T_{\bh}<0)} 
\sum_{i,j=1}^n {\bef}_{i,\bT \rightarrow  j, k \mathbf h} = \ 
\sum_{\bT}^{(T_{\bh}<0)} 
\sum_{i,j=1}^n {\bef}_{i, \bT - \bT \rightarrow j, k \bh -\bT} = \
\sum_{\bT}^{(T_{\bh}<0)} 
\sum_{i,j=1}^n {\bef}_{i,\mathbf 0\rightarrow j, k \bh -\bT} \nonumber \\
&=& \
\sum_{\bT}^{(T_{\bh}>0)} 
\sum_{i,j=1}^n {\bef}_{i,\mathbf 0\rightarrow j, k \bh +\bT} = \
\sum_{\bT}^{(T_{\bh}>k)} 
\sum_{i,j=1}^n {\bef}_{i,\mathbf 0\rightarrow j,\bT}.
\end{eqnarray}
Then the net force, by the $L_{\bh}$ part, acting on the whole half cell bar $B_{\bh}$ is
\begin{equation}
{\bF}_{\bh} = \ 
\sum_{i = 0}^ \infty {\bF}_{{\bh}, i} = \ 
\sum_{\bT}^{(T_{\bh}<0)} 
\sum_{i,j=1}^n T_{\bh}{\bef}_{i,\mathbf 0\rightarrow j,\bT} = \ 
\sum_{\bT}^{(T_{\bh}>0)} 
\sum_{i,j=1}^n T_{\bh}{\bef}_{i,\mathbf 0\rightarrow j,\bT} = \
\frac 12\ \sum_{\bT\neq 0} \sum_{i,j=1}^n T_{\bh}{\bef}_{i,{\mathbf 0}\rightarrow j,\bT}.\label{b08}
\end{equation}

On the other hand, the total potential energy of the MD cell is 
\begin{equation}
E_{p,MD} = \sum_{i>j=1}^n \varphi^{(2)} (\left| {\br}_i-{\br}_j\right| ) + 
{\frac 1 2} \sum_{\bT\neq 0} \sum_{i,j=1}^n\varphi^{(2)} (\left| \br_i-\br_j-\bT\right| ), \label{b09}
\end{equation}
where half of potential energy between any two distinct cells is assigned to each.
Let us introduce a main interaction tensor (dyad):
\begin{equation}
\Tepmain = -\frac1\Omega \left[\left( \frac{\partial E_{p,MD}}{\partial\ba}\right) {\ba} + \left( \frac{\partial E_{p,MD}}{\partial\bb}\right) {\bb} +\left( \frac{\partial E_{p,MD}}{\partial\bc}\right) {\bc} 
\right].\label{b10.0}
\end{equation}
It follows that
\begin{equation}
\Tepmain = \frac 1{2\Omega}\ \sum_{\bT\neq 0}\sum_{i,j=1}^n\ {\bef}_{i,\mathbf 0\rightarrow j,\bT}{\bT}.\label{b10}
\end{equation}
Remembering  $\bh^\prime\cdot\sh=\delta_{\bh,\bh^\prime}\Omega$ 
($\bh^\prime=\ba$, $\bb$, or $\bc$), 
we have 
\begin{equation}
\bF_{\bh} = \Tepmain\cdot \sh.\label{b11}
\end{equation}

Using Eq.(\ref{b12}), the left hand side of Eq.(\ref{b07}) becomes
\begin{equation}
\frac 1{N_{\bh}} M_R\ddot {\br}_{RC} = 
\frac 1{N_{\bh}} \sum_{\bT\in R_{\bh}}
\sum_{i=1}^nm_i\left(\ddot\br_i + \ddot\bT\right)  = 
\frac{M_{cell}} {N_{\bh}} \sum_{\bT\in R_{\bh}} \ddot \bT,  \label{b13}
\end{equation}
where the total cell mass is $M_{cell}=\sum_{i=1}^nm_i$. Noticing that 
$\ddot\bT=T_{\ba}\ddot\ba +T_{\bb}\ddot\bb +T_{\bc}\ddot\bc$,
Eq. (\ref{b13}) may be written as
\begin{equation}
\frac 1{N_{\bh}} M_R\ddot {\br}_{RC} = 
\alpha _{\bh,\ba}\ddot\ba +
\alpha _{\bh,\bb}\ddot\bb +
\alpha _{\bh,\bc}\ddot\bc,\label{b19}
\end{equation}
where 
\begin{equation}
\alpha _{\mathbf{h,h}^\prime}=
\frac{M_{cell}}{N_\bh}
\sum_{\bT\in R_\bh}
T_{{\bh}^\prime}\ \ (\bh^\prime = \ba,\ \bb,\ \bc).\label{b20}
\end{equation}
In the $R_\bh$ part,  $T_\bh$ is always non-negative, but for any  $T_{\bh^\prime\neq\bh}$, we can assume that there exists another 
$-T_{\bh^\prime}$ that cancels it in the above summation. We therefore can neglect all non-diagonal terms $\alpha _{\bh,\bh^\prime\neq\bh}$. 
Then Eq.(\ref{b07}) becomes 
\begin{equation}
\alpha _{\bh,\bh} \ddot\bh = 
\left( \Tepmain + \Tup\right ) \cdot \sh\ \ 
(\bh = \ba,\ \bb,\ \bc).\label{b21}
\end{equation}
The dynamical equation Eq.(\ref{b21}) is essentially the same as in our previous works\cite{lww, lwssc}, where  only constant external pressure was considered.

\section{Statistics of instantaneous dynamics}\label{sec:Statistics}

It would be reasonable to replace the main interaction tensor $\Tepmain$ in 
Eq.(\ref{b21}) by the whole internal stress tensor, including kinetic-energy and interaction terms\cite{haile,irving,meclellan,swenson,tsai,tadmor}, 
so that the period vectors are driven by the imbalance between the internal and external stresses. 
In this section we are going to pursue this goal. 
Let us focus on the interaction term based on statistics of  indistinguishable translated states first and move on to the kinetic-energy one by considering forces associated with transport of momentum in the following subsections.

\subsection{Full interaction term in internal stress}

\begin{figure}
  \begin{center}
    \hspace{-0.3cm}
    \includegraphics[width=0.7\textwidth]{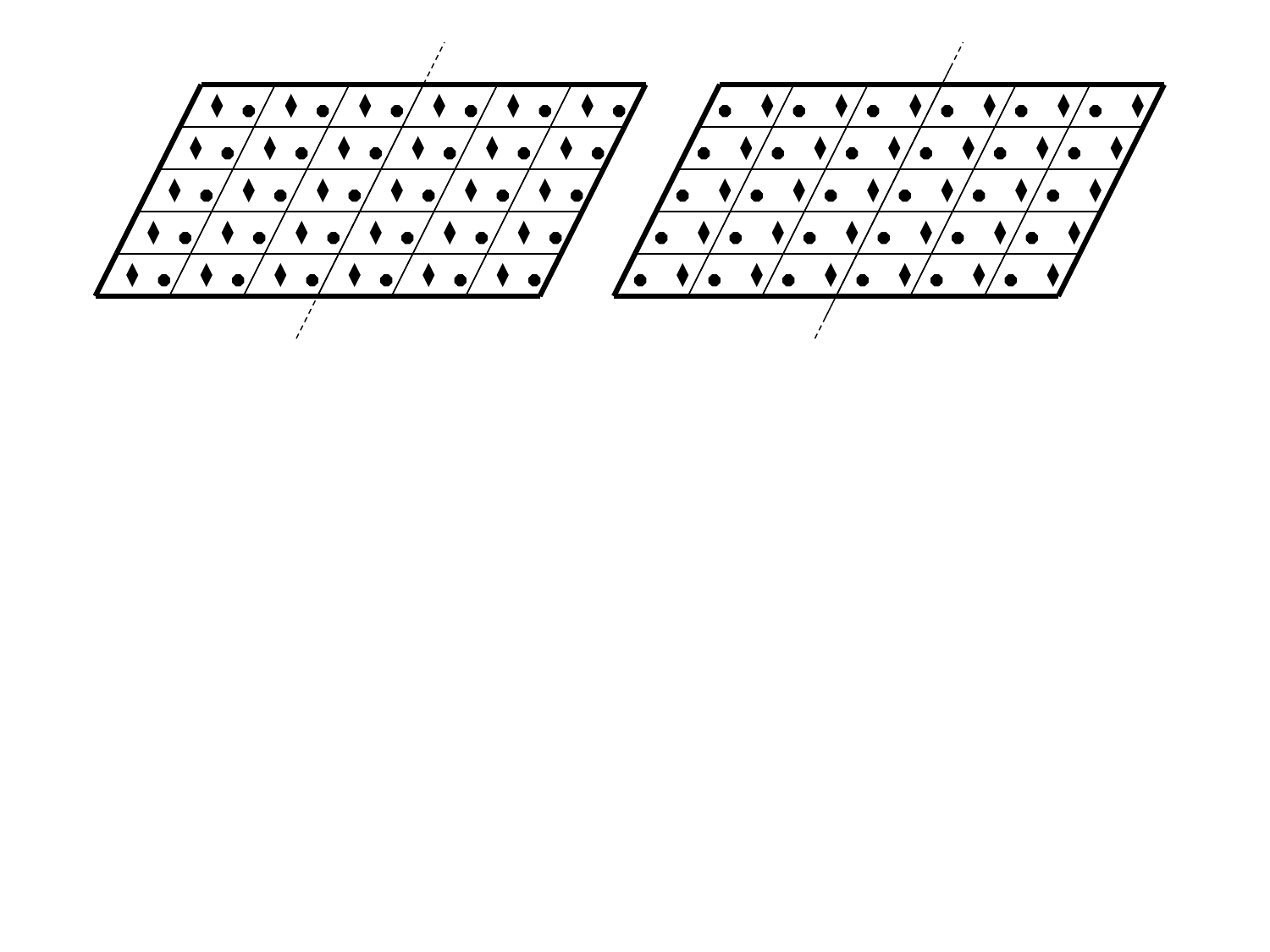}
  \end{center}
  \vspace{-9.4cm} \hspace{0.9cm} $P$
  \vspace{-0.0cm} \hspace{5.2cm} $P$ \\
  \vspace{2.2cm} \hspace{-1.5cm} $P^{\prime}$
  \vspace{-0.0cm} \hspace{5.2cm} $P^{\prime}$
  \vspace{7.3cm} 
  \vspace{-1.7cm} 
  \vspace{-5.5cm}
  \caption{
A sketch for two distinct states of the system 
which are exactly the same in all microscopic details 
except being translated slightly 
as a whole relative to each other.
As a set of image particles, the black diamonds are 
on the right side in each cell of the right state, but  
on the left side in each cell of the left state.
The black disks as another set of image particles are 
on the other sides in the states. 
Please also note that in these states all right parts from plane $PP^{\prime }$ 
have the same number of cells. 
}
  \label{fig2}
\end{figure}

In Appendix B.3 of his book\cite{haile}, Haile derives the full interaction term of internal stress as
\begin{equation}
\Tep=\frac 1V\sum_{I>J}\left(\bef_{J\rightarrow I}\right) \left(\br_I-\br_J\right), \label{i01}
\end{equation}
where the particle indices $I$ and $J$ are running over all the material and $V$ is the total volume of the material, by taking an unweighted average of internal interactions passing through all possible parallel cross sections of the material. Following this idea, we may take an average of Eq.(\ref{b07}) over different locations of the cutting plane $PP^{\prime }$ in Fig. 1. However the result is just a combination of  Eqs.(\ref{b07}) and (\ref{b03}).

Still, there are some situations that should be included for statistics. For instance, as in Fig. 2, the two distinct states of the system are exactly the same in all microscopic details except being translated slightly as a whole relative to each other. Please note that in these states all right parts from plane $PP^\prime$ have the same number of cells. Since they cannot be distinguished from a macroscopic point of view, we should 
take an unweighted average of  Eq.(\ref{b07}) over all such configurations. Among these, Eq. (\ref{b07}) is the same except the net force $\bF_{\bh}$, acting on the half cell bar $B_{\bh}$ by the $L_{\bh}$ part, as in Fig. 1. Equivalently, this means that the 
net force $\bF_{\bh}$ in Eq. (\ref{b07}) is replaced by the unweighted average of it over all possible parallel locations of cutting planes $PP^\prime$ that pass through the MD cell in Fig. 1. For clarity, let us still use  $PP^{\prime }$ in its original meaning (i.e., no ``running''), but employ $QQ^{\prime }$ for such a running plane. Let us do the statistics in the following four cases.

When the plane $QQ^\prime$ runs from left to right passing through the MD cell, the probability for MD particle $i$ appearing on the left side of 
$QQ^\prime$ is $\left(\bh-\left(\br_i-\br_0\right)\right) \cdot\sh/\Omega$, where 
$\br_0$ is the position vector of the left-bottom and far-away vertex of the MD cell. It can also be 
written as $\left(\bhz-\br_i\right) \cdot\sh/\Omega$ with $\bhz=\bh+\br_0$. 
The corresponding averaged net force acting on this particle by the $L_{{\mathbf h}}$ part, namely 
\begin{equation}
\bef_{\bh,c1,i} = \frac {\left(\bhz-\br_i\right) \cdot \sh} {\Omega}
\sum_{\bT}^{(T_{\bh}<0)}\sum_{j=1}^n  \bef_{j,\bT\rightarrow i,\mathbf 0},
\end{equation}
should be substracted from $\bF_{\bh}$. 

Let us call all cells 
$\bT=T_{\ba}\ba+T_{\bb}\bb+T_{\bc}\bc$ with $T_{\bh}=0$ as 
$T_{\bh}=0$ slab. Then the averaged net force, by all particles 
in the $T_{\bh}=0$ slab of the left side of $QQ^\prime$, 
acting on the  
half cell bar $B_{\bh}$ of the right side of $QQ^\prime$ (see Fig. 1), should be added to $\bF_{\bh}$. 

Now let us consider the case of such interactions by the $T_{\bh}=0$ slab on all cells $\bh$, $2\bh$, $3\bh$, $4\bh$, etc., till the surface, only.  
The resulting averaged net force should be equal to the averaged one by MD cell on 
all $\bT$ cells with $T_{\bh}>0$, for the periodicity of the system, 
\begin{equation}
\bef_{\bh,c2} =
\sum_{\bT}^{(T_{\bh}>0)}\sum_{i,j=1}^n  
 \frac {\left(\bhz-\br_i\right) \cdot \sh} {\Omega}
\bef_{i,\mathbf 0\rightarrow j,\bT}=
\frac {1} {\Omega}
\sum_{\bT}^{(T_{\bh}>0)}\sum_{i,j=1}^n  \left( \bef_{j,\bT\rightarrow i,\mathbf 0} \right) \left(\br_i-\bhz\right) \cdot \sh.
\end{equation}

The third case is about such interactions by the MD particles of the left side of $QQ^\prime$ on the MD particles of the right side. The probability for any pair of MD particles $i$ and $j$ being on different sides of $QQ^\prime$ is 
$\left| \left(\br_i-\br_j\right) \cdot \sh\right|/\Omega$. Then all the averaged interaction of pairs is 
\begin{eqnarray}
\bef_{\bh,c3}
&=& \frac 1\Omega \sum_{i>j}^n \left(\bef_{j,\mathbf 0\rightarrow i,\mathbf 0}\right) \left(\br_i-\br_j\right) \cdot \sh\nonumber \\
&=& \frac 1{2\Omega} \sum_{i\neq j}^n \left(\bef_{j,\mathbf 0\rightarrow i,\mathbf 0}\right) \left(\br_i-\bhz-\br_j+\bhz\right) \cdot \sh\nonumber \\
&=& \frac 1{2\Omega} \sum_{i\neq j}^n \left(\bef_{j,\mathbf 0\rightarrow i,\mathbf 0}\right) \left(\br_i-\bhz\right) \cdot \sh
+ \frac 1{2\Omega} \sum_{i\neq j}^n \left(\bef_{i,\mathbf 0\rightarrow j,\mathbf 0}\right) \left(\br_j-\bhz\right) \cdot \sh\nonumber \\
&=& \frac 1\Omega \sum_{i\neq j}^n \left(\bef_{j,\mathbf 0\rightarrow i,\mathbf 0}\right) \left(\br_i-\bhz\right) \cdot \sh .
\end{eqnarray}

The final case is about the rest of such interactions. 
They are by all particles of the $T_{\bh}=0$ slab of the left side of plane $QQ^\prime$ but excluding MD particles, acting on the MD particles of the right side of $QQ^\prime$. Let particle $i$ be in the MD cell ($\bT=0$) and particle $j$ in another cell ($\bT\neq 0$) of the $T_{\bh}=0$ slab.
The probability for these particles being on opposite sides of the plane $QQ^\prime$ is
$\left| \left(\br_i-\br_j-\bT\right) \cdot \sh\right|/\Omega = \left| \left(\br_i-\br_j\right) \cdot \sh\right|/\Omega$, 
since $T_{\bh}=0$. 
It is zero for $i=j$. 
Then the averaged net force is
\begin{eqnarray}
{\mathbf f}_{{\mathbf h},c4}
&=&\frac 1{\Omega} \sum_{\bT\neq 0}^{(T_{\bh}=0)} \sum_{i>j}^n \left(\bef_{j,\bT\rightarrow i,\mathbf 0}\right) \left(\br_i-\br_j\right) \cdot \sh\nonumber \\
&=&\frac 1{2\Omega} \sum_{\bT\neq 0}^{(T_{\bh}=0)} \sum_{i,j=1}^n \left(\bef_{j,\bT\rightarrow i,\mathbf 0}\right) \left(\br_i-\br_j\right) \cdot \sh\nonumber \\
&=&\frac 1{2\Omega} \sum_{\bT\neq 0}^{(T_{\bh}=0)} \sum_{i,j=1}^n \left(\bef_{j,\bT\rightarrow i,\mathbf 0}\right) \left(\br_i-\bhz-\br_j+\bhz\right) \cdot \sh\nonumber \\
&=&\frac 1{2\Omega} \sum_{\bT\neq 0}^{(T_{\bh}=0)} \sum_{i,j=1}^n \left(\bef_{j,\bT\rightarrow i,\mathbf 0}\right) \left(\br_i-\bhz\right) \cdot \sh
      -\frac 1{2\Omega} \sum_{\bT\neq 0}^{(T_{\bh}=0)} \sum_{i,j=1}^n \left(\bef_{j,\bT\rightarrow i,\mathbf 0}\right) \left(\br_j-\bhz\right) \cdot \sh\nonumber \\
&=&\frac 1{2\Omega} \sum_{\bT\neq 0}^{(T_{\bh}=0)} \sum_{i,j=1}^n \left(\bef_{j,\bT\rightarrow i,\mathbf 0}\right) \left(\br_i-\bhz\right) \cdot \sh
      -\frac 1{2\Omega} \sum_{\bT\neq 0}^{(T_{\bh}=0)} \sum_{i,j=1}^n \left(\bef_{j,\bT - \bT \rightarrow i,-\bT}\right) \left(\br_j-\bhz\right) \cdot \sh\nonumber \\
&=&\frac 1{2\Omega} \sum_{\bT\neq 0}^{(T_{\bh}=0)} \sum_{i,j=1}^n \left(\bef_{j,\bT\rightarrow i,\mathbf 0}\right) \left(\br_i-\bhz\right) \cdot \sh
      +\frac 1{2\Omega} \sum_{\bT\neq 0}^{(T_{\bh}=0)} \sum_{i,j=1}^n \left(\bef_{i,-\bT\rightarrow j,\mathbf 0}\right) \left(\br_j-\bhz\right) \cdot \sh\nonumber \\
&=&\frac 1{\Omega} \sum_{\bT\neq 0}^{(T_{\bh}=0)} \sum_{i,j=1}^n \left(\bef_{j,\bT\rightarrow i,\mathbf 0}\right) \left(\br_i-\bhz\right) \cdot \sh.
\end{eqnarray}

Then the average net force $\overline\bF_{\bh}^\prime$, acting on the half cell bar $B_{\bh}$ by the $L_{\bh}$ part in Fig. 1 is
\begin{eqnarray}
\overline\bF_{\bh}^\prime 
&=& \bF_{\bh} - \left(\sum_{i=1}^n\bef_{\bh,c1,i}\right) +\bef_{\bh,c2} + \bef_{\bh,c4} + \bef_{\bh,c3} \nonumber \\
&=& \bF_{\bh} + \frac 1\Omega \sum_{\bT\neq 0} \sum_{i,j=1}^n \left(\bef_{j,\bT\rightarrow i,\mathbf 0}\right) \left(\br_i-\bhz\right)\cdot \sh 
+ \frac 1\Omega \sum_{i\neq j}^n \left(\bef_{j,\mathbf 0\rightarrow i,\mathbf 0}\right) \left(\br_i-\bhz\right)\cdot\sh\nonumber \\
&=& \bF_{\bh}+\frac 1\Omega \sum_{i=1}^n \bF_i\br_i\cdot\sh - \frac 1\Omega \left(\sum_{i=1}^n\bF_i \right)\bhz\cdot\sh\nonumber \\
&=& \bF_{\bh} + \frac 1\Omega \sum_{i=1}^n \bF_i\br_i\cdot \sh,\end{eqnarray}
where we have used Eq. (\ref{b12}). Now, let us introduce
\begin{equation}
\Tep_p=\frac 1\Omega \sum_{i=1}^n\bF_i\br_i. \label{i04.2}
\end{equation}
Then we have
\begin{equation}
\overline\bF_{\bh}^\prime = \bF_{\bh} +\Tep_p\cdot\sh=\left(\Tepmain + \Tep_p\right)\cdot\sh=\Tep\cdot\sh,\label{newi04}
\end{equation}
where the full interaction term of the internal stress is
\begin{equation}
\Tep=\Tepmain+\Tep_p.\label{i04}
\end{equation}
As a matter of fact, Eq.(\ref{i04}) is equivalent to Eq.(\ref{i01}). If all MD particles are in an equilibrium state, i.e. $\bF_i=0$, for $i=1,2,...,n$, the full interaction term of the internal stress becomes
$\Tep=\Tepmain$,  which is normally \textit{not} zero.
%The above statistics changes Eq.(\ref{b21}) into
%\begin{equation}
%\alpha _{\bh,\bh} \ddot\bh = \left( \Tep +\Tup\right ) \cdot \sh\ \ (\bh=\ba, \bb, \bc).\label{nb21}
%\end{equation}

\subsection{Kinetic-energy term in internal stress}

In Appendix B.2 of his book \cite{haile}, Haile derived the kinetic-energy term of internal stress by considering a force associated with pure transport of momentum across geometrical planes,  without collisions or any other interactions. Newton's second law tells us that for a specific particle, any force acting on it causes its momentum to change, and any change in momentum must  be done by a force or forces acting on it. Let us consider particle $p$ with mass $m > 0$ running with a fixed velocity $\mathbf v\neq 0$ without being acted on by any force. However, if we accept the idea of forces associated with pure transport of momentum, when the particle passes through any geometrical plane, such a force becomes manifest. The questions are where the force comes 
from and under what conditions it should be considered.

In order to answer the questions, we need to distinguish between material-based and space-based systems\cite{malvern}. When we talk about some fixed specific particles, we are employing material-based systems, which always contain the same particles. In this description, Newton's second law can definitely be satisfied and no addtitional forces are needed. On the other hand, when we talk about a particle passing through a geometrical plane, we may be considering space-based systems.

Let us view the above uniformly moving particle  $p$ in a space-based system description. Imagine $A$ and $B$ are two neighboring empty space boxes that are separated by a plane $PP^\prime$. The single particle $p$ passes through this plane from $A$ into $B$ during a short time interval $\triangle t >0$. Let us define the spatial system $S_A$ as the system containing all particles in box $A$, and similarly the spatial system $S_B$ of any particles in box $B$. In these systems, the total number of particles may change with time. The momentum increase for system $S_B$ during $\triangle t$ is $m\mathbf v$. In order to satisfy Newton's second law applied on system $S_B$, there must be a force $\mbf=m\mathbf v /\triangle t \neq 0$, acting on system $S_B$ during this period. Where does this force come from? Obviously, there is no regular force. 
If we do the same thing for system $S_A$, we find that another force $\mbf^\prime=-\mbf\neq 0$ is needed to satisfy Newton's second law on system $S_A$ for the same period $\triangle t$. These two additional forces are  needed at the same time only when particle $p$ passes from one into the other system. So we may regard them as the interaction forces between the two systems, which means that $S_A$ acts on $S_B$ with the force $\mbf$ and $S_B$ acts on $S_A$ with the other force $\mbf^\prime$. This interaction obeys Newton's Third Law. If every aspect remains the same except that particle $p$ moves from $B$ into $A$ with velocity $-{\mathbf v}$, the forces between the two systems remain exactly the same.

In the material-based system description, there is no transport of momentum into or from the system and therefore no need to consider any additional forces.  If the space-based system description is employed, the net rate of transport of momentum into a system from a neighboring system, $\dot {{\mathbf M}\text{,}}$ should be added to the acceleration side of Newton's second law. The corresponding additional force $\mbF=\dot\mathbf M$ is needed and must be added to the other side of the equation to keep it satisfied. The neighboring space-based  system losing $\dot\mathbf M$ needs another force 
$-\mbF$ to satisfy its own Newtonian dynamics. These additional forces can be regarded as interactions between the space-based systems, satisfying Newton's Third Law. 
They do not change the physics, but are necessary to maintain the same physics when the point of view is changed. 

As a matter of fact, space-based system descriptions are widely used in study of continuous media, gas, and other systems with ``changing mass'', as materials or particles are very difficult  to trace in such situations. For example, an ideal gas is placed into a sealed cuboid container in a dynamical equilibrium state. If we imagine to use a fixed geometrical plane $PP^\prime$  to cut the gas into a left and a right part, then both parts are space-based systems, and particles may move from one into the other. Then the force associated with transport of momentum acrossing plane $PP^\prime$ must be considered and should be balanced by the regular force of 
collisions between the gas and the container in the surface, so that each of the two parts is in an equilibrium state.

For a periodic system, if we use a material-based  system description, we need to trace particles and also have to explicitely consider their collisions with the external wall when they reach the bulk surface. Instead, let us employ the space-based system description to focus on what happens inside the bulk, with all that is happening at the surface considered in the form of the external stress. So specifically, the above $L_{\bh}$ and $R_{\bh}$ parts in Fig. 1 are both space-based systems. Now let us consider the above statistics over the indistinguishable translated states
with the help of $QQ^\prime$ again, but of the force associated with transport of momentum. 
If we say the total amount of such indistinguishable translated states is the cell volume $\Omega$, the amount of them where MD particle $i$ can cross plane $QQ^\prime$ during unit time is $\left| \dot\br_i \cdot \sh\right|$, with momentum $m_i \dot\br_i$ being carried each. 
Although, during this period of time, the period vectors may change, the change should be very limited compared with the priod vectors themselves especially when approaching equilibrium states, then neglected. 
Then the averaged rate of momentum change  
\begin{equation}
\bef_{\bh,tm}=\frac 1\Omega \sum_{i=1}^n m_i\dot \br_i \dot \br_i \cdot \sh
\end{equation}
should be added to $\bF_{\bh}$. As a result, Eq.(\ref{newi04}) is updated to 
\begin{equation}
\overline\bF_{\bh}=\overline\bF_{\bh}^\prime + \bef_{\bh,tm} = \left(\Tep + \Ttau^\prime\right)\cdot\sh,\label{newi05}
\end{equation}
where the instantaneous kinetic-energy term of the internal stress is
\begin{equation}
\Ttau^\prime=\frac 1\Omega \sum_{i=1}^n m_i\dot \br_i \dot \br_i .
\label{newi06}
\end{equation}
Defining the instantaneous internal stress (or virial stress in \cite{tadmor}) as 
\begin{equation}
\Tpi^\prime=\Tep+\Ttau^\prime,  \label{newc09.2}
\end{equation}
the period dynamics Eq.(\ref{b21}) becomes 
\begin{equation}
\alpha _{\bh,\bh} \ddot\bh=\left( \Tpi^\prime + \Tup\right ) \cdot \sh\ \ (\bh=\ba, \bb, \bc).\label{newb22}
\end{equation}

The observable period vectors showing fixed values under certain external conditions (e.g. constant external pressure and temperature) 
should not depend on the directions of particles' motions. 
Now let us perform a further unweighted average of 
Eq.(\ref{newb22}) over all moving directions of the MD particles.  We first arrive at 
the averaged kinetic-energy term of internal stress by averaging Eq.(\ref{newi06}) : 
\begin{equation}
\Ttau=\frac 1{3\Omega}\sum_{i=1}^n m_i\left| \dot \br_i\right| ^2\TI=\frac 2{3\Omega}E_{k,MD}\TI,
\end{equation}
where $E_{k,MD}$ is the total kinetic-energy of the MD particles. 
The forces corresponding to this part of the internal stress should be balanced by the part of the external forces involved in collisions between the particles in the bulk surface and the surrounding external walls, as in the above example of an ideal gas. Accordingly, the averaged internal stress from Eq.(\ref{newc09.2}) is
\begin{equation}
\Tpi=\Tep+\Ttau.\label{newc09.22}
\end{equation}
Then the period dynamics Eq.(\ref{newb22}) changes into
\begin{equation}
\alpha _{\bh,\bh} \ddot\bh = \left( \Tpi +\Tup\right ) \cdot \sh\ \ (\bh=\ba, \bb, \bc).\label{nb222}
\end{equation}

\section{Summary and Discussion}\label{sec:Summary}

Keeping Newton's second law for MD particles and applying it to macroscopic half-systems with addtional statistics 
over indistinguishable translated states and forces associated with transport of momentum applied, 
we arrived at a group of coupled dynamical equations, Eqs. (\ref{b03}) and (\ref{nb222}), for periodic systems under constant external stress. Eq.(\ref{nb222}) shows that the system period vectors are driven by the imbalance between the internal and external stresses. 

The internal stress has both kinetic-energy and interaction terms. The kinetic-energy term was obtained from the statistics of forces associated with transport of momentum when the two halves of the system are recognized as space-based ones. A further statistics over movement directions of particles was carried out based on the observability of the period vectors. This effect should be reflected in collisions  
of surface particles with external walls. 

For further discussion of the total kinetic-energy of the MD particles, 
let us divide all MD particles into some local groups and suppose each group $g$ only shows a 
macroscopic motion of velocity $ \dot \mathbf{r}_g $, while every particle $i$ in the group has  
a thermal motion of velocity $ \dot  \mathbf{u}_i $ and the 
macroscopic motion of velocity $ \dot \mathbf{r}_g $. In other words, for any particle $i$ of the group, 
the total velocity is  $ \dot \mathbf{r}_i= \dot \mathbf{u}_i  + \dot \mathbf{r}_g$, and 
the total thermal motion of the group is zero, i.e.
\begin{equation}
\sum_{i \in g} m_i \dot  \mathbf{u}_i =0.  \label{bfmmt12}
\end{equation}
Then the kinetic-energy for the group
\begin{equation}
E_{k,g}=\frac 1{2}\sum_{i\in g} m_i\left(\dot \mathbf{u}_i  + \dot \mathbf{r}_g \right) ^2
=\frac 1{2}\sum_{i\in g} m_i \dot \mathbf{u}_i  ^2 +
 \frac 1{2}\sum_{i\in g} m_i \dot \mathbf{r}_g  ^2,
\end{equation}
%=\frac 1{2}\sum_{i\in g} m_i \dot \mathbf{u}_i  ^2 +
% \left( \sum_{i\in g} m_i \dot \mathbf{u}_i  \right) \cdot \dot \mathbf{r}_g  +
% \frac 1{2}\sum_{i\in g} m_i \dot \mathbf{r}_g  ^2
which means the thermal kinetic energy and the macroscopic kinetic energy of the group is seperated. 

Then the total kinetic-energy of the MD particles $E_{k,MD}$ can also be written as a sum of the total thermal kinetic energy 
and the total macroscopic kinetic energy of the MD particles. The former reflects the temperature of the system. The latter 
may be zero for many systems, e.g. crystals, and may be significantly important in other systems, e.g. fluids. 
Considering Eqs. (\ref{bmmt12}) and (\ref{bfmmt12}), the total macroscopic momentum of the MD cell should be  zero, i.e.
\begin{equation}
\sum_{g\in\text{MD}}\sum_{i\in g}m_i\dot{{\mathbf r}}_g=0. 
\end{equation}

The full interaction term is devided into the main interaction term (Eq.(\ref{b10.0})) and the rest (Eq.(\ref{i04.2})). The former reflects the main interactions between the two halves of the system, while the latter covers the statistical change of such interactions over the indistinguishable translated states. Similar to Eq. (\ref{b10.0}),  the full interaction term, Eq. (\ref{i04}), may also be written as
\begin{equation}
\Tep=-\frac 1\Omega \sum_{\mathbf z\in\text{DOF}} \left( \frac{\partial E_{p,MD}}{\partial\mathbf z}\right)\mathbf z,
\label{verynew}
\end{equation}
where DOF refers to all degrees of freedom of the system. With this form, the dynamics may be easily extended to cases of more accurate forces, e.g. forces from quantum mechanical computations.

As defined in Eq.(\ref{b20}), the mass-like coefficient $\alpha _{\bh,\bh}$ is determined by the size and shape of the macroscopic bulk and only affects the rate of change for period vectors. Since we are only interested in properties of the inner part of the bulk, we therefore simply suggest to set 
%\begin{equation}
$\alpha _{\ba,\ba}=\alpha _{\bb,\bb}=\alpha _{\bc,\bc}=M_{cell}$, 
%\end{equation}
in order to make computation feasible.
%Then Eq.(\ref{nb222}) becomes 
%\begin{equation}
%M_{cell} \ddot\bh = \left( \Tpi +\Tup\right ) \cdot \sh\ \ (\bh=\ba, \bb, \bc).
%\end{equation}

With simulations for a series values of constant external stress, this approach may also be used in studying system properties that change with external stress. For instance,  the very interesting   
piezoelectric and piezomagnetic effects could be simulated.

%\newpage

\section*{Acknowledgements}
The author wishes to thank Prof. Ding-Sheng Wang, Institute of Physics, Prof. Si-Yuan Zhang, Changchun Institute of Applied Chemistry, Prof. S. S. Wu, Jilin University, Prof. En-Ge Wang, Beijing University, P.R. China, Dr. Kenneth Edgecombe, Dr. Hartmut Schmider, Dr. Malcolm J. Stott, Dr. Kevin Robbie, Queen's University, Canada, and Dr. Xiaohua Wu, the Royal Military College of Canada, for their helpful discussions and earnest encouragement.

%\newpage

%\newpage

%\newpage

\end{document}